\documentclass[twocolumn,showpacs,pre]{revtex4}
\usepackage{dcolumn}
\usepackage{graphicx}
\usepackage{graphicx,amssymb,amsmath,revsymb}
\usepackage{natbib}

\newcommand{\toy}{\kappa}		

\begin{document}

\title{Non--local fluctuation correlations in active gels}

\author{D. A. Head$^{1,2}$ and D. Mizuno$^{3}$}

\affiliation{$^{1}$Institute of Industrial Science, University of Tokyo, Meguro-ku, Tokyo 153-8505, Japan}
\affiliation{$^{2}$Institut f\"ur Festk\"orperforschung, Theorie II, Forschungszentrum J\"ulich 52425, Germany}
\affiliation{$^{3}$Institute for the advanced study, Kyushu Univ., 812-8581, Fukuoka, Japan}

\date{\today}

\begin{abstract}
Many active materials and biological systems are driven far from equilibrium by embedded agents that spontaneously generate forces and distort the surrounding material. Probing and characterizing these athermal fluctuations is essential for understanding the properties and behaviors of such systems. Here we present a mathematical procedure to estimate the local action of force-generating agents from the observed fluctuating displacement fields. The active agents are modeled as oriented force dipoles or isotropic compression foci, and the matrix on which they act is assumed to be either a compressible elastic continuum or a coupled network-solvent system.  Correlations at a single point and between points separated by an arbitrary distance are obtained, giving a total of three independent fluctuation modes that can be tested with microrheology experiments. Since oriented dipoles and isotropic compression foci give different contributions to these fluctuation modes, ratiometric analysis allows us characterize the force generators. We also predict and experimentally find a high-frequency ballistic regime, arising from individual force generating events in the form of the slow build-up of stress followed by rapid but finite decay.  Finally, we provide a quantitative statistical model to estimate the mean filament tension from these athermal fluctuations, which leads to stiffening of active networks.   
\end{abstract}


\maketitle

\section{Introduction}

Living organisms are inherently non--equilibrium systems that continually harvest external resources in order to maintain their activity and complex hierarchical ordering~\cite{AlbertsBook,SchrodingerBook}. Thermodynamic equilibrium is therefore expected to apply, at best, to length and time regimes that are not associated with biological processes. The cellular cytoskeleton, a complex assembly composed of filamentous proteins, motor proteins and other accessories which provides a cellular scaffolding~\cite{HowardBook,BoalBook,BrayBook,Paluch2006,Kolomeisky2007}, is a case in point. This dynamic structure hydrolyses energy transfer molecules such as adenosine tri\-phosphate (ATP) to promote morphological and mechanical variations, thus regulating many biomechanical processes such as organelle transport, muscle contraction, and cell division~\cite{BoalBook,BrayBook,Paluch2006,Kolomeisky2007}. Since such processes need not obey the statistics of thermodynamic equilibrium~\cite{Lau2003,Lau2009}, novel analytic techniques need to be devised both to predict the viscoelastic response, and to physically describe the active entities from experimental data.

The violation of thermodynamic equilibrium has been demonstrated in model {\em in vitro} cytoskeletons consisting of an actin filament network coupled to the motor protein myosin~\cite{Mizuno2007}, which generates $\sim$ pN forces upon ATP hydrolysis~\cite{HowardBook,Kolomeisky2007}. The material's mechanical viscoelastic response was simultaneously determined by two microrheology protocols, one {\em (passive)} that assumes thermodynamic equilibrium and a second {\em (active)} that does not~\cite{Gittes1997,Levine2000,Crocker2000,Levine2001,Wilson2009,Mizuno2009}. Any difference between these two protocols, which has been observed at low ({\em i.e.} biological) frequencies, is attributed to the existence of non--equilibrium fluctuations~\cite{Mizuno2007,Mizuno2008,Mizuno2009}. A departure from Gaussian fluctuation statistics has also been observed in active, entangled actin--microtubule mixtures~\cite{Brangwynne2008}. The existence of athermal force generators in various model systems has been shown to spontaneously create non--equilibrium flow patterns and structures~\cite{Nedelec1997,Surrey2001,Backouche2006}, requiring novel analytical techniques to understand their collective behavior ~\cite{Liverpool2003,Kruse2004,Liverpool2005,Kruse2005,Liverpool2006,Voituriez2006,Ruhle2008}. Introducing permanent crosslinkers such as biotin/avidin complexes into {\em in vitro} cytoskeletal systems blocks filament flow and reduces complexity to that of fluctuations in a static medium. This allows for a physical characterization of the athermal fluctuations, leading the way to a fundamental understanding of their biological implications.

Here we derive the spatial and temporal correlations in the fluctuating displacement field generated by a population of athermal force generators, or `firers', dispersed in a linear viscoelastic material, in a form that can be directly compared to microrheology data. Firers are modeled as force dipoles, generating stress and strain fields that propagate long distances as dictated by material response. As a consequence the correlations induced by motor activity are {\em non--local}; that is, a single motor activation generates correlated motion in distant parts of the medium. For the surrounding environment, we consider two material types, {\em (i)}~a single--component elastic continuum, and {\em (ii)}~a two--component system consisting of a sparse elastic network that is frictionally coupled to an interspersed, incompressible solvent, known as the `2--fluid model'~\cite{Levine2000,Onuki1997,Tanaka2000,Furukawa2004,Tanaka2006}. The latter is closer in spirit to the cellular cytoskeleton, but as our calculations will show, the consequence of the explicit solvent--network description to predicted microrheology data is negligibly small in most cases.  Thus the single--component model should approximately apply to systems of interest, as has been already assumed in prior microrheology studies. Since we expect a critical role of strong local compression induced by molecular force dipoles on the mechanics of networks, we do not assume material incompressibility, in contrast to previous approaches~\cite{Lau2003,Lau2009}. Macroscopic isotropy and homogeneity, relevant to {\em in vitro} experiments, are assumed throughout ~\cite{Bursac2005,Alamo2008,Duits2009,Brawley2009}.

For both classes of material, we derive exact expressions for the displacement power spectrum $C_{ij}(\omega)$, which is the magnitude of displacement fluctuations $\langle u^{2}\rangle$ expressed as a function of frequency~$\omega$, and the two--point spectrum $D_{ij}({\bf R},\omega)$ that quantifies correlations in displacements between points separated by a vector distance~${\bf R}$. These quantities can be directly compared to data from 1 and 2--point microrheology experiments, respectively. We find 3 sources of variation with frequency: {\em (i)}~The power spectrum of the forces generated by individual firers. For an actin--myosin system, in which the contractile motor stresses build up gradually but are rapidly released, the power spectrum is expected to take the form $\sim \omega^{-2}$, as has been observed~\cite{Mizuno2008,MacKintosh2008,Brangwynne2008,Levine2009}. In reality the time required to release the developed compression is short but finite, so to probe this effect we solve a simple model with a controllable release time, and predict a crossover to a high--frequency ballistic regime $\sim \omega^{-4}$. Experimental data confirming the existence of this ballistic regime is also provided. {\em (ii)}~The frequency dependence of the mechanical response (shear modulus~$\mu(\omega)$) of the matrix. The linear mechanical response of an actin gel has been studied intensively and is now relatively well understood as a network of semi--flexible polymers~\cite{MacKintosh1995, Isambert1996,Kroy1996,Morse1998,Gittes1998,Tassieri2008}, at least for affine deformation~\cite{Head2003a,Head2003b,Wilhelm2003,Gardel2004,Liu2007,Buxton2007,Das2007,Broedersz2009} and negligible contribution from crosslinkers or other accessory proteins~\cite{Lieleg2008,Ward2008,Kasza2009}. {\em (iii)}~When the material is assumed to obey the 2--fluid model, a third source of variation comes from the frictional coupling between the network and solvent, which becomes tighter at higher frequencies and suppresses compressional/longitudinal response modes. For parameters relevant to actin--myosin systems, our calculations indicate that this contribution is expected to be small.

\begin{figure}
\center{\includegraphics[width=8.5cm]{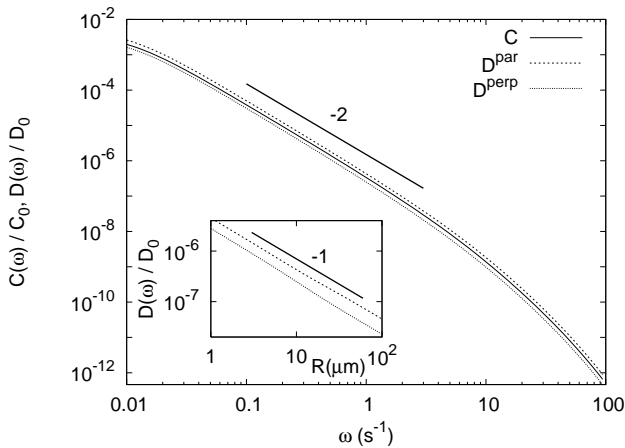}}
\caption{Example of the one--point and two--point correlation functions $C_{ij}(\omega)$ and $D_{ij}(R,\omega)$, resp., from the analytical expressions (\ref{e:Cij}) and~(\ref{e:Dij_result}) for the two--fluid model. The inset shows $D_{ij}(\omega=1s^{-1})$ for different separations~$R$. Following~\cite{Levine2009}, the network shear modulus is $\mu(\omega)=\mu_{0}(1+[-i \omega/\omega_{0}]^{3/4})$ with $\omega_{0}=10$ s$^{-1}$. For the dipole moment power spectrum $|\kappa(\omega)|^{2}$, we used the finite--release model (\ref{e:PSD}) with $\tau=100$s and $\alpha=10^{-3}$. Other parameters are $\nu=0.3$ and $\Gamma=10^{11}$ kg m$^{-3}$ s$^{-1}$ corresponding to a mesh size $\xi\approx100$nm and solvent viscosity $\eta\approx10^{-3}$ Pa s. The normalization factors are $C_{0}=\bar{c}\kappa_{0}^{2}/\mu_{0}^{2}a$ and $D_{0}=\bar{c}\kappa_{0}^{2}/\mu_{0}^{2}R$ with $\kappa_{0}^{2}=2\tau\langle\kappa\rangle^{2}$.}
\label{f:example}
\end{figure}

An example of the correlation functions for the 2--fluid model from our calculations is given in Fig.~\ref{f:example}, where we have projected out the parallel $D^{\parallel}(R)$ and perpendicular $D^{\perp}(R)$ components of $D_{ij}({\bf R})$ in the usual way, as \mbox{$D_{ij}({\bf R})=D^{\parallel}(R)\hat{R}_{i}\hat{R}_{j}+D^{\perp}(R)(\delta_{ij}-\hat{R}_{i}\hat{R}_{j})$} with $\delta_{ij}$ the Kr\"onecker delta, $R=|{\bf R}|$ and $\hat{\bf R}={\bf R}/R$. 
These calculations are based on the assumption of a population of randomly distributed and oriented, independently--firing active agents modeled as force dipoles. The inset shows the spatial variation of $D_{ij}({\bf R})$ with $R$ (as $R^{-1}$) which is consistent with experiments and specific to force dipoles; higher--order multipoles give different power laws, as explained in Sec.~\ref{s:Dij} below. The scaling with $R$ does not determine the symmetry of the underlying force dipole, and we identify two cases of relevance to actin--myosin systems, namely an {\em oriented} force dipole consisting of two nearby placed equal--and--opposite forces, corresponding to a single myosin minifilament attached to the actin network, and an isotropic {\em compression center} corresponding to the orientational average of the first class.  This latter case is expected to arise due to fluctuations in the local concentration of motor protein, or filament rupture leading to local network compression.  Below we derive $C_{ij}$ and $D_{ij}$ for both classes of dipole, explicitly confirming $D_{ij}\sim 1/R$ but revealing qualitatively different low--frequency behavior for the parallel correlations~$D^{\parallel}(R)$, which approaches $2D^{\perp}(R)$ for oriented dipoles in the incompressible limit, but entirely vanishes for compression centers. Hence, by investigating the length--scale dependence of each athermal fluctuation mode or by comparing the amplitude of different independent modes, it is now possible to identify the class of firer in a system of unknown composition, and further test the validity of the theoretical modeling or fine--tune the model to specific experimental systems.

Before proceeding, it should be noted that active gels are not the only situation where internally--generated force impulses are of relevance. Soft glasses and colloidal gels relaxing towards their equilibrium state exhibit local spontaneous rearrangement events or {\em micro--collapses}~\cite{Cipelletti2000,Bouchaud2001,Bouchaud2002}, generating dipolar force perturbations much as in active gels. Elastic propagation of stress from local irreversible rearrangements is the essence of various physical descriptions of glassy systems (see~\cite{Picard2004,Picard2005,Goyon2008,Manning2007} and references therein), which are also typically modeled as a consequence of creation and annihilation of local force dipoles. The results obtained in this study thus apply to a broader range of non-equilibrium systems than suggested by the primary, biophysical motivation.  

This paper is arranged as follows. In Sec.~\ref{s:single_firer} we define both oriented force dipoles and isotropic compression centers in terms of a frequency--dependent dipole moment~$\kappa(\omega)$. We show that a simple model with slow stress build--up followed by a rapid but finite--duration release exhibits a high--frequency ballistic regime in its power spectrum, $|\kappa(\omega)|^{2}\sim\omega^{-4}$. Microrheology data consistent with this prediction is also given. The correlation functions of the athermal fluctuating displacements fields are derived in Sec.~\ref{s:correlation_functions}. We first define the mechanical response of 2 types of materials, single-- and two--component, in Sec.~\ref{s:materials}, and then derive $C_{ij}(\omega)$ and $D_{ij}({\bf R},\omega)$ for oriented force dipoles for both materials in Sec.~\ref{s:Cij} and Sec.~\ref{s:Dij} respectively. In Sec.~\ref{s:isotropic} the same quantities are derived for isotropic compression centers. Finally in Sec.~\ref{s:discussion} we estimate the mean network tension from the magnitude of athermal fluctuations in an active gel.

%
%

\section{Individual firers}
\label{s:single_firer}

An active agent or `firer' here refers to any local construct embedded within a continuum body that is capable of spontaneously generating mechanical stress. For example, in an actin--myosin system each firer is a bipolar minifilament of ${\mathcal O}(10)$ myosin motors that consumes the local supply of ATP upon activation~\cite{Humphrey2002,Surrey2001,Kiehart1986}. We consider only the linear response, and leave non--linear considerations to future work (such as the mixed response mode introduced in~\cite{Mizuno2009}). The active forces can then be written as a multipole expansion of the form $\sim f\varepsilon^{n}\nabla^{n}\delta({\bf r})$, where $\varepsilon$, $f$ are microscopic length and force scales, resp., and $n=1,2\ldots$ corresponds to dipoles, quadrupoles {\em etc}. The monopole term $n=0$ vanishes for a stable network as the total force must balance, but there is no {\em a priori} reason to neglect higher terms. Thus firers are here taken as dipolar to leading order, and we shall later confirm that experiments are in line with this assumption.

\begin{figure}
\center{\includegraphics[width=8.5cm]{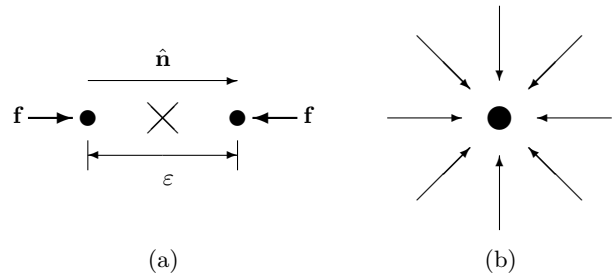}}
\caption{(a)~A single contractile force dipole of size $\varepsilon$, force $f$ and orientation $\hat{\bf n}$. (b)~Schematic representation of an isotropic compression center as the orientational average of~(a).}
\label{f:schematic}\end{figure}

A single firer in an active gel at the position ${\bf x}$ is modeled as a pair of equal--and--opposite forces $\pm f\hat{\bf n}$ located at ${\bf x}\mp\frac{\varepsilon}{2}\hat{\bf n}$, respectively, where $\hat{\bf n}$ is the firer orientation, as shown in Fig.~\ref{f:schematic}(a). For simplicity we take the limit $\varepsilon\rightarrow0$ and $f\rightarrow\infty$ with a fixed dipole moment $\kappa=\varepsilon f$. This reduces each firer to a point force dipole of zero spatial extent, which is a valid approximation when addressing quantities that vary on length scales much larger than the size of the firer (as in the typical microrheology situation). The force density due to a single point dipole is

\begin{equation}
{\bf f}^{\rm ani}({\bf r},t)=\kappa(t)\hat{\bf n}(\hat{\bf n}\cdot\nabla)\delta({\bf r}-{\bf x})
\label{e:f_dip}
\end{equation}

\noindent{}where the suffix `ani' differentiates these dipoles, with orientation $\hat{\bf n}$, from the isotropic compression centers defined below.  The time--dependent strength of the dipole moment $\kappa(t)$, which as defined is positive for contractile firers and negative for expansive ones, denotes a train of activation events, including the possibility of `dead--time' between pulses with $\kappa\equiv0$

As mentioned in the introduction, we also consider isotropic compression centers as in Fig.~\ref{f:schematic}(b), applicable when local clusters of coupled firers activate near--simultaneously or when independent firers are locally aggregated. Formally these are the average of (\ref{e:f_dip}) over the orientation vector $\hat{\bf n}$,

\begin{equation}
{\bf f}^{\rm iso}({\bf r},t)=\frac{1}{3}\kappa(t)\nabla\delta({\bf r}-{\bf x})
\label{e:f_iso_dip}\end{equation}
\noindent{}where we have used the identity $\hat{n}_{i}\hat{n}_{j}\rightarrow\frac{1}{3}\delta_{ij}$ under averaging. Such isotropic force dipoles or dilation/compression centers have been studied in a number of contexts~\cite{Siems1968,Schwarz2002}. 


\subsection{Power spectrum for a single firer}
\label{s:finite_release}

\begin{figure}
\center{\includegraphics[width=8.5cm]{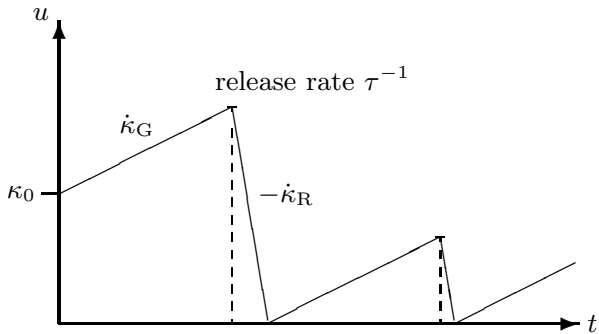}}
\caption{Notation for the finite--release model. The dipole moment $\toy$ increases linearly at a rate $\dot{\toy}_{\rm G}$, initiates release at a rate $\tau^{-1}$ after which it decreases linearly at a rate $-\dot{\toy}_{\rm R}$ until reaching zero when the cycle restarts.}
\label{f:finite_release}
\end{figure}

The temporal dependence of the dipole moment $\kappa(t)$ can by itself significantly contribute to the system's fluctuation spectrum. It is experimentally observed that strains induced by myosin minifilaments slowly build up over time scales that in some cases extend up to $\sim$1---10~s, and then rapidly release~\cite{Mizuno2007}, allowing the internally generated tension to relax to an unstrained state. This rapid release generates a pseudo--diffusive power spectrum $\sim\omega^{-2}$ for frequencies larger than the inverse characteristic build--up time, despite the lack of any diffusing field~\cite{MacKintosh2008,Levine2009}. 
At frequencies far below this time, a plateau $\sim\omega^{0}$ is expected.  However, although rapid, the time for stress release is nonetheless finite. As already discussed in~\cite{Mizuno2007}, stress build--up is accompanied by significant local network compression deriving from the non--linear response of actin filaments. In order to relax, it is therefore necessary for the network to dilate, the rate of which is limited by the inflow of solvent. A relaxation time of $\sim0.1$~s was estimated by balancing frictional and elastic forces during relaxation, and found to be consistent with experimental observation~\cite{Mizuno2007}.
To quantify the effect of a finite release time on the power spectrum, we consider here a simple model in which the dipole moment of a single firer $\kappa(t)$ slowly increases in time before stochastically releasing stress as shown in Fig.~\ref{f:finite_release}. The model is defined by 3 parameters: The rate of increase of the moment $\toy$ in the growth phase, $\dot{\toy}_{\rm G}>0$; the corresponding rate in the release phase, $-\dot{\toy}_{\rm R}<0$, and the release rate $\tau^{-1}$ at which growth switches to release. Note that $\toy$ always returns to zero before growth restarts, so it is bounded and does not diffuse. The steady state solution is found following the procedure described in Appendix~\ref{app:finite_release}. The resulting power spectrum $|\kappa(\omega)|^{2}$ due to this single firer is

\begin{eqnarray}
|\kappa(\omega)|^{2}
&=&
\frac{2\tau\langle\toy\rangle^{2}}{1-\alpha}
\left\{
\frac{1}{1+\omega^{2}\tau^{2}}
-
\frac{\alpha^{2}}{1+\alpha^{2}\omega^{2}\tau^2}
\right\}
\label{e:PSD}
\end{eqnarray}

\begin{figure}
\center{\includegraphics[width=8.5cm]{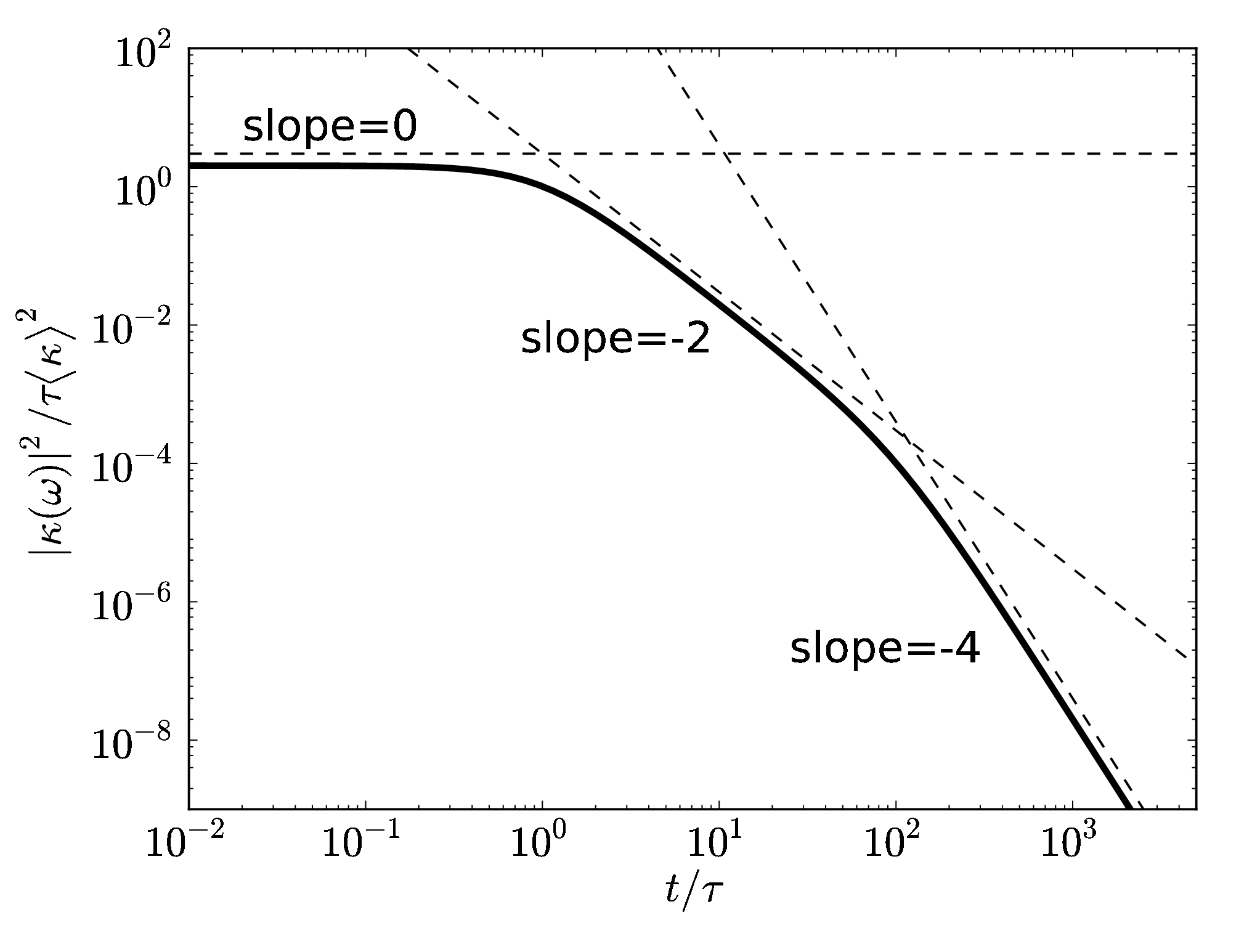}}
\caption{The thick line gives the power spectrum $|\kappa(\omega)|^{2}$ of a single dipole  according to~(\ref{e:PSD}), with $\alpha=\dot{\toy}_{\rm G}/\dot{\toy}_{\rm R}=1/100$, so the mean release time is 1\% of the growth time. The straight dashed lines have given slopes and both axes have been made dimensionless by suitable normalization.}
\label{f:finite_release_example}
\end{figure}

\noindent{}where $\alpha=\dot{\toy}_{\rm G}/\dot{\toy}_{\rm R}<1$ is the ratio of growth to release rates, and $\langle\toy\rangle=\tau\dot{\toy}_{\rm G}$ is the mean moment in steady state. This expression admits 3 regimes: {\em (i)}~A low frequency plateau $|\kappa(\omega)|^{2}\sim\omega^{0}$ for $\omega\ll\tau^{-1}$; {\em (ii)}~An intermediate pseudo--diffusive regime $|\kappa(\omega)|^{2}\sim\omega^{-2}$ for $\tau^{-1}\ll\omega\ll(\alpha\tau)^{-1}$, where the upper crossover frequency is the inverse mean release time; and {\em (iii)}~A high frequency ballistic regime $|\kappa(\omega)|^{2}\sim\omega^{-4}$ for $\omega\gg(\alpha\tau)^{-1}$. In the rapid release limit $\alpha\rightarrow0$, the ballistic regime vanishes and the pseudo--diffusive regime extends to all high frequencies, as expected. An example for $\alpha=1/100$ is given in Fig.~\ref{f:finite_release_example}.

For an actin--myosin system, the ballistic regime should be clearly resolved for frequencies greater than $\sim$10~Hz using the estimated release time $\sim0.1$~s discussed above. This is around the frequency where thermal fluctuations start to dominate~\cite{Mizuno2007,Mizuno2009}, so achieving a satisfactory signal--to--noise ratio is already challenging. Furthermore this frequency is above the elastic plateau for typical network densities, so the ballistic spectrum may be obscured by additional frequency variation coming from the network viscoelasticity. For highly crosslinked networks, however, thermal noise is suppressed and the elastic plateau extends to higher frequencies~\cite{Gittes1998}, so the contribution of the fluctuation of firers, $|\kappa(\omega)|^{2}$, should dominate the frequency dependence of the  experimentally--measured displacement power spectrum, even at high frequencies.

\begin{figure}
\center{\includegraphics[width=8.5cm]{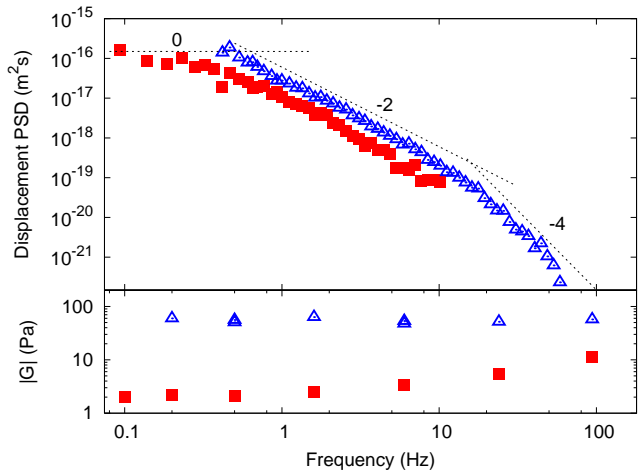}}
\caption{ {\em (Color online)} The upper panel shows the athermal component of the displacement power spectral density for loosely {\em (filled squares)} and a tightly {\em (open triangles)} crosslinked networks obtained from microrheology~\cite{Mizuno2009}. The straight lines have the given slopes. The lower panel shows the absolute shear moduli $|G(\omega)|$ for the same networks, demonstrating plateau response and confirming the observed frequency--dependence in the power spectrum is not due to network viscoelasticity.}
\label{f:Exp_Plot}
\end{figure}

Experimental confirmation of the ballistic $\omega^{-4}$ regime is presented in Fig.~\ref{f:Exp_Plot}, where the athermal component of the power spectral density (PSD) of both loosely and tightly crosslinked networks are given, as measured using the same microrheology protocol as detailed in~\cite{Mizuno2009}. Also shown are absolute shear moduli $|G|$ of both networks over the same frequency range, confirming that the tightly crosslinked network is still in its elastic plateau up to frequencies of $\sim100$Hz. Thus the downturn in the PSD is not due to the network response. It is however consistent with the $\sim\omega^{-4}$ ballistic regime predicted above, and we conclude this is its origin. Note that while there is evidence of a $\sim\omega^{0}$ regime at low frequencies for the sparsely crosslinked network, the data here is also consistent with a weak frequency variation, possibly due to the very slow, irreversible changes in network structure in response to strong
force generations.

%
%

%
%

%
%
\section{Athermal fluctuation correlations}
\label{s:correlation_functions}

In this section, correlations in the fluctuations of the displacement field generated by a population of many firers are derived, for two classes of material. The starting point for the calculations is the displacement field around a single active firer, which will here be denoted by the material--dependent propagator ${\bf P}({\bf x},t;\hat{\bf n})$ that gives the displacement at a point ${\bf x}$ and time $t$ due to the activity of a firer at ${\bf x}={\bf 0}$ and an earlier time $t=0$. The total displacement ${\bf u}$ at a point ${\bf r}$ and time $t$ can then be found by linear superposition of all active firers,

\begin{equation}
{\bf u}({\bf r},t)
=
\int {\rm d}{\bf x}
\int {\rm d}s
\int {\rm d}\hat{\bf n}
\:
{\bf P}({\bf r}-{\bf x},t-s;\hat{\bf n})
c({\bf x},s;\hat{\bf n})
\label{e:summed}
\end{equation}

\noindent{}where $c({\bf x},s;\hat{\bf n})$ describes the number of firers per unit volume of orientation $\hat{\bf n}$ at time $s$. We assume activity is statistically uniform and uncorrelated in both space and time, with a mean firing rate per unit volume (of any orientation) denoted $\bar{c}$. Note that in an actin-myosin system, $c$ refers to myosin mini--filaments and not simply the concentration of myosin.

%
%
\subsection{Mechanics of materials}
\label{s:materials}

To proceed it is necessary to specify ~${\bf P}({\bf x},t;\hat{\bf n})$ by describing the mechanical response of the material. Two materials are considered here, both represented on the continuum level under the assumption that the quantities of interest vary on lengths larger than the coarse--graining length of the network. First we consider the two--fluid model that has already been used to calculate the viscoelastic response of filament networks in solution~\cite{Levine2000,MacKintosh2008,Levine2009}. In brief, the system is modeled as a coupled two--component system, consisting of a fixed isotropic elastic network interacting {\em via} local friction with an incompressible fluid. The network displacement ${\bf u}$ and solvent velocity ${\bf v}$ obey

\begin{eqnarray}
0&=&
\mu \nabla^{2}{\bf u}+(\mu+\lambda)\nabla(\nabla\cdot{\bf u})
+\Gamma({\bf v}-\partial_{t}{\bf u})
+{\bf f}^{\rm ext}
\nonumber\\
0&=&
\eta\nabla^{2}{\bf v}-\nabla P-\Gamma({\bf v}-\partial_{t}{\bf u})
\label{e:two_fluid}
\end{eqnarray}

\noindent{}with $\mu$, $\lambda$ the elastic Lam\'e coefficients and $\eta$ the solvent viscosity. The pressure gradient $\nabla P$ imposes \mbox{$\nabla\cdot{\bf v}=0$} and the external force ${\bf f}^{\rm ext}$ describes all active force generators. The frictional coupling coefficient $\Gamma$ is estimated to be $\sim \eta/\xi^{2}$ with $\xi$ the network mesh size~\cite{Levine2000}. The solution is found by transforming to Fourier space, {\em i.e.}

\begin{equation}
c({\bf q},\omega;\hat{\bf n})=\int{\rm d}{\bf x}\int{\rm d}t\:{\rm e}^{i(\omega t-{\bf q}\cdot{\bf x})}c({\bf x},t;\hat{\bf n})
\end{equation}

\noindent{}and projecting the transformed ${\bf u}$ and ${\bf v}$ into components perpendicular and parallel to $\hat{\bf q}$, solvent incompressibility requiring that the latter vanishes for ${\bf v}$. The transformed propagator ${\bf P}({\bf q},\omega)$ can be written in terms of the projected components of a single firer ${\bf f}^{\rm act}({\bf q},\omega)$ as

\begin{eqnarray}
{\bf P}({\bf q},\omega)
&=&
\frac{ [{\bf f}^{\rm act}\cdot{\bf \hat{q}}]\hat{\bf q} }{B(\omega)q^{2}-i\omega\Gamma}
+
\frac{ {\bf f}^{\rm act}-[{\bf f}^{\rm act}\cdot{\bf \hat{q}}]\hat{\bf q} } {q^{2}G(\omega)}
\label{e:propagator}
\end{eqnarray}

\noindent{}where $B=2\mu+\lambda$ and $G=\mu-i\omega\eta$ is the total shear modulus. The force ${\bf f}^{\rm act}({\bf q},\omega)$ is the transform of either the oriented dipole~(\ref{e:f_dip}), or the isotropic compression~(\ref{e:f_iso_dip}),

\begin{eqnarray}
{\bf f}^{\rm act}({\bf q},\omega)
&=&
\left\{
\begin{array}{ccl}
i\kappa(\omega)q[\hat{\bf q}\cdot\hat{\bf n}]\hat{\bf n} & : & \mbox{oriented}
\\
\frac{1}{3}i\kappa(\omega){\bf q} & : & \mbox{isotropic}
\end{array}
\right.
\label{e:f_dip_choice}
\end{eqnarray}

\noindent{}The expression (\ref{e:propagator}) is valid for $\Gamma\gg\eta q^{2}$, {\em i.e.} $q\ll\xi^{-1}$, which is anyway necessary to justify the use of the continuum equations~\cite{Levine2009}. The Poisson ratio $\nu=\lambda/[2(\mu+\lambda)]$ can be estimated by assuming an affine network deformation, giving the frequency--independent value $\nu=1/4$, as explained in Appendix~\ref{app:poisson_ratio}.

The second material considered here is a classical elastic body. Formally this corresponds to the $\Gamma\equiv0$ limit in~(\ref{e:two_fluid}), but allows the calculations to be carried out in real (rather than Fourier) space, which makes a small difference to the calculation for single point fluctuations as explained below. These calculations start from the known result for the displacement field due to a force monopole, from which the dipole field is found and the displacement correlations calculated by integrating the products of displacements. Details are given in Appendix~\ref{app:elasticity_calculations}.

%
%
\subsection{Single--point athermal fluctuations}
\label{s:Cij}

Fluctuations in the network strain field can either be measured at the same point, or at points separated by a finite distance, corresponding to 1 and 2--point microrheology, respectively. While fluctuations in active materials have both a thermal and an athermal origin, they may often be treated as independent and extracted separately as described in~\cite{Mizuno2007}. We calculate here the athermal contributions to these quantities deriving solely from firers, beginning with the 1--particle correlation function $C_{ij}(t)=\langle u_{i}({\bf 0},0)u_{j}({\bf 0},t)\rangle$, where the angled brackets denote averaging over space, time and dipole orientations $\hat{\bf n}$ (isotropic compression centers are considered in Sec.~\ref{s:isotropic}). By making use of~(\ref{e:summed}) and assuming uncorrelated, uniform firers, $C_{ij}(t)$ can be rewritten in frequency space as

\begin{equation}
C_{ij}(\omega)
=
\frac{\bar{c}}{24\pi^{3}}
\delta_{ij}
\left\langle\int d{\bf q}\left|{\bf P}({\bf q},\omega)\right|^{2}\right\rangle_{\hat{\bf n}}
\label{e:Cij_def}
\end{equation}

\noindent{}where $\langle\cdots\rangle_{\hat{\bf n}}$ denotes averaging over orientation $\hat{\bf n}$ only.

For the propagators (\ref{e:propagator}) and~(\ref{e:f_dip_choice}), there is a short--wavelength divergence in~(\ref{e:Cij_def}) at high~$q$ resulting from the slow decay of~${\bf P}$, $P^{2}\sim q^{-2}$. The physical origin of this singularity is the diverging magnitude of displacements induced at positions arbitrarily close to an active firer. Recall however that the expression for ${\bf P}$ in~(\ref{e:propagator})  corresponds to {\em point} dipoles; for real finite--size dipoles, ${\bf P}$ will become strongly attenuated at length scales corresponding to the dipole size and this divergence will be avoided. In microrheology, a further small distance cut-off appears from the finite radius $a$ of the probe particle, which excludes dipoles from a range $a$ of the origin. We focus on this form of cut--off and introduce a maximum magnitude for the wave vector, $q^{\rm max}=\pi/2a$~\cite{Levine2000}, and perform the integration in (\ref{e:Cij_def}) using ${\bf P}$ defined by the oriented dipole in~(\ref{e:f_dip_choice}), giving

\begin{eqnarray}
C_{ij}(\omega)
&=&
\frac{1}{60\pi a}\bar{c}\kappa^{2}
\delta_{ij}
\left\{
\frac{2}{3|G|^{2}}
+
\frac{1}{|B|^{2}}
\psi\left(2a/\pi\ell\right)
\right\}
\nonumber\\
\psi(z)
&=&
1-\frac{1}{\Re(B)}
\Re\left\{Bz\sqrt{-i}\arctan\left[
\frac{1}{z\sqrt{-i}}
\right]\right\}
\nonumber\\
\label{e:Cij}
\end{eqnarray}

\noindent{}where the length $\ell=\sqrt{B/\omega\Gamma}$ is discussed below, $\Re()$ denotes the real part and the same root of $\sqrt{-i}$ should be chosen for both occurrences. Note that $\kappa^{2}=|\kappa(\omega)|^{2}$ is the power spectrum for a single firer, and there is implicit frequency dependence in both $B(\omega)$ and $G(\omega)$. A representative example of $C_{ij}(\omega)$ for an actin--myosin system is given in Fig.~\ref{f:example}. The corresponding result from the real--space elasticity calculations is

\begin{equation}
C_{ij}^{\rm el}
=
\frac{3-7\nu+5\nu^{2}}{(1-\nu)^2}
\frac{\bar{c}\kappa^{2}}{180\pi\mu^{2}}
\frac{1}{a}
\delta_{ij}
\label{e:Cij_elastic}
\end{equation}

\noindent{}Note that this differs slightly from the low--frequency limit of (\ref{e:Cij}), $C_{ij}(\omega=0)=\frac{2\frac{3}{4}-7\nu+5\nu^{2}}{(1-\nu)^2}\frac{\bar{c}\kappa^{2}}{180\pi\mu^{2}}\frac{1}{a}\delta_{ij}$, by the constant coefficient in the numerator polynomial, which is attributable to the different manner of handling the short wavelength cut--off in real and Fourier space~\cite{Furukawa2004}.

\begin{figure}
\center{\includegraphics[width=8cm]{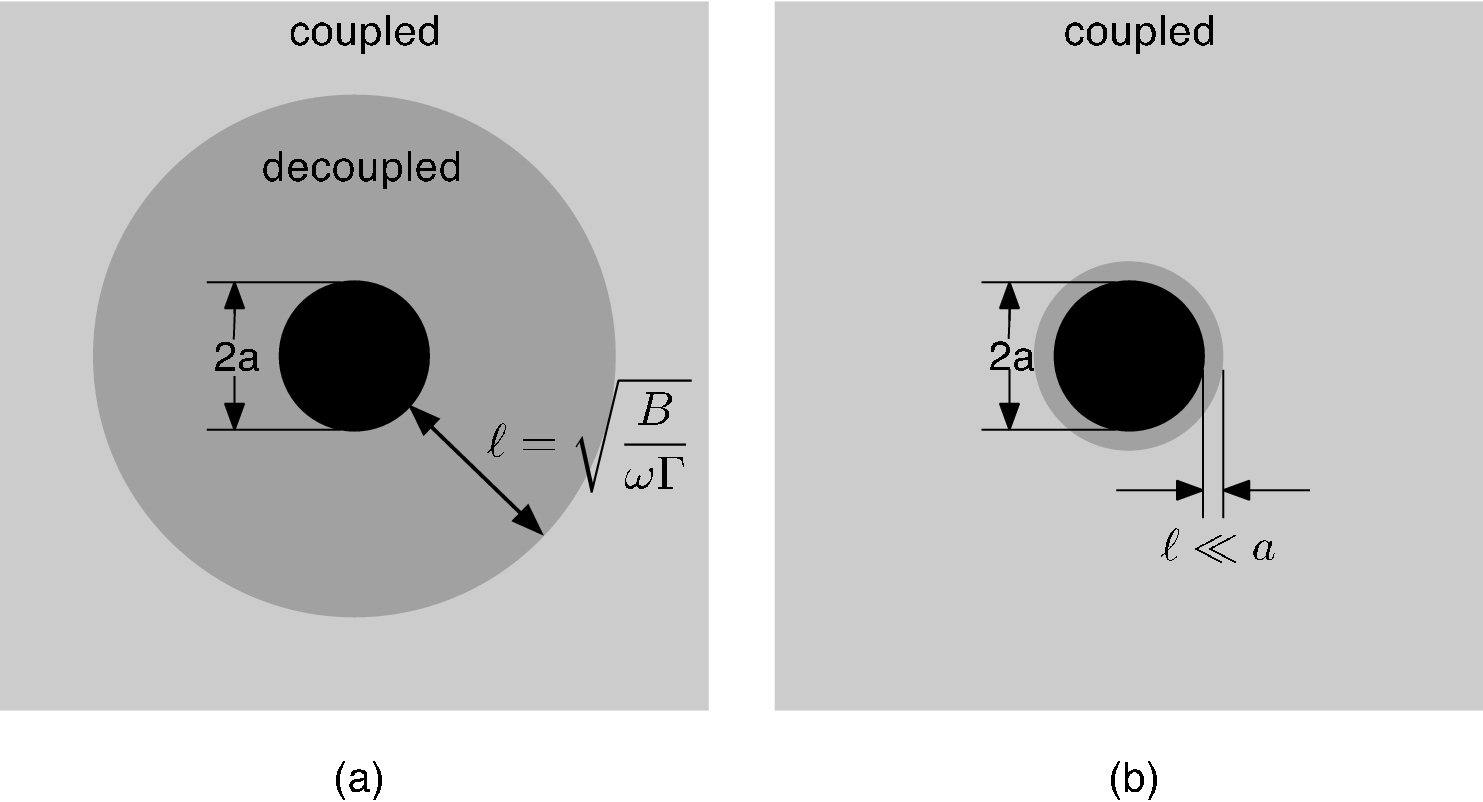}}
\caption{Schematic representation of the effects of the coupling between the network and the solvent on the 1--particle response function~$C_{ij}(t)=\langle u_{i}({\bf 0},0)u_{j}({\bf 0},t)\rangle$. The crossover between the coupled response at large lengths {\em (light grey)} and the decoupled response at short lengths {\em (dark grey)} is given by the frequency--dependent penetration length $\ell=\sqrt{B/\omega\Gamma}$, and the dominant response determined by the ratio of $\ell$ to the probe radius~$a$. (a)~For low frequencies, $\ell/a\gg1$ and $C_{ij}$ is dominated by decoupled network and solvent responses. (b)~For high frequencies, $\ell\ll a$ and $C_{ij}$ is now dominated by the coupled network--solvent response.}
\label{f:1pt_coupling_schematic}
\end{figure}

Equation~(\ref{e:Cij}) is composed of shear/transverse modes $\propto|G|^{-2}$, and compressional/longitudinal modes~$\propto|B|^{-2}$. The contribution from transverse modes is independent of the network--solvent coupling, which is entirely contained in the longitudinal contribution. The magnitude of this term depends on the ratio of the cut--off radius $a$ to the frequency--dependent {\em penetration length} $\ell=\sqrt{B/\omega\Gamma}$, which determines the range of the longitudinal network modes~\cite{Levine2009}. Two different coupling regimes occur in the limits $a\ll\ell$ and $a\gg\ell$, as schematically shown in Fig.~\ref{f:1pt_coupling_schematic}. For low frequencies such that $a\ll\ell$, the network is fully decoupled from the solvent and $\psi(z)\approx1$. In this limit the longitudinal response is dominated by the elastic network. In the opposite limit $a\gg\ell$, network and solvent are fully coupled and now $\psi(z)\approx|B|^{2}\pi^{4}/(80\omega^{2}\Gamma^{2}a^{4})$. This term therefore decays as $\sim\omega^{-2}$ at high frequencies, so the longitudinal response vanishes and only the shear contributions remain.

%
%

%
%
\subsection{Two--point athermal correlations}
\label{s:Dij}

Spatial correlations of athermal fluctuations are encoded in the 2--particle function $D_{ij}({\bf R},t)=\langle u_{i}({\bf 0},0)u_{j}({\bf R},t)\rangle$, which quantifies correlated motion over a vector distance~${\bf R}$. Using~(\ref{e:summed}) and again assuming spatio--temporally uniform activity, this can be rewritten in reciprocal space as

\begin{equation}
D_{ij}({\bf q},\omega)
=
\bar{c}\left\langle P_{i}({\bf q},\omega;\hat{\bf n})P_{j}(-{\bf q},-\omega;\hat{\bf n})\right\rangle_{\hat{\bf n}}
\label{e:Dij_defn}
\end{equation}

\noindent{}This can be evaluated following the procedure outlined in Appendix~\ref{app:Dij}. In terms of the projections parallel and perpendicular to the line of centers ${\bf R}$, 
\mbox{$D_{ij}({\bf R})=D^{\parallel}(R)\hat{R}_{i}\hat{R}_{j}+D^{\perp}(R)(\delta_{ij}-\hat{R}_{i}\hat{R}_{j})$} with \mbox{$R=|{\bf R}|$}, we find

\begin{eqnarray}
D^{\parallel}(R,\omega)
&=
\displaystyle{\frac{\bar{c}\kappa^{2}}{60\pi R}}
&
\left\{
\frac{1}{|G|^{2}}
+
\frac{3}{|B|^{2}}\psi^{\parallel}(R/\ell)
\right\}
\nonumber\\
D^{\perp}(R,\omega)
&=
\displaystyle{\frac{\bar{c}\kappa^{2}}{120\pi R}}
&
\left\{
\frac{1}{|G|^{2}}
+
\frac{3}{|B|^{2}}
\psi^{\perp}(R/\ell)
\right\}
\label{e:Dij_result}
\end{eqnarray}

\noindent{}where $\psi^{\perp}(z)=\Re\{2iB[1-(1+z\sqrt{-i})e^{-z\sqrt{-i}})]/z^{2}\}/\Re(B)$ and  $\psi^{\parallel}(z)=\Re\{{Be^{-z\sqrt{-i}}}\}/\Re(B)-\psi^{\perp}(z)$, and the root is chosen such that $\Re({z\sqrt{-i}})>0$. See Fig.~\ref{f:example} for an example relevant for actin--myosin systems. For comparison, the real--space calculations give

\begin{eqnarray}
D^{\rm el}_{\parallel}(R)
&=&
\frac{\bar{c}\kappa^{2}}{60\pi\mu^{2}}
\frac{1}{R}
\nonumber\\
D^{\rm el}_{\perp}(R)
&=&
\frac{16\nu^{2}-20\nu+7}{(1-\nu)^{2}}
\frac{\bar{c}\kappa^{2}}{480\pi\mu^{2}}
\frac{1}{R}
\label{e:Dij_elastic}
\end{eqnarray}

\noindent{}which is exactly the zero--frequency limit of~(\ref{e:Dij_result}).

\begin{figure}
\center{\includegraphics[width=8cm]{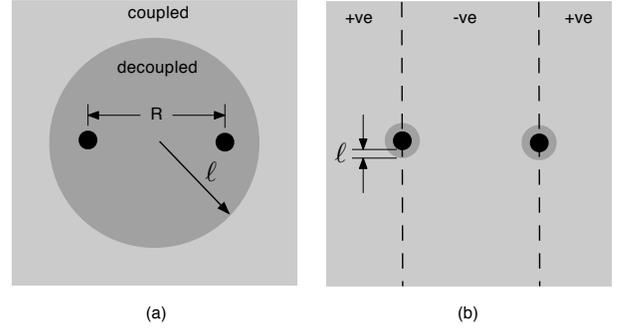}}
\caption{Same as Fig.~\ref{f:1pt_coupling_schematic} but for the 2--particle correlation function~$D_{ij}({\bf R},t)=\langle u_{i}({\bf 0},0)u_{j}({\bf R},t)\rangle$. The crossover between the decoupled~(a) and coupled~(b) regimes is determined by the inter--particle spacing~$R$ relative to the penetration length~$\ell$. Also shown in (b) are the regions responsible for negative and positive correlations to $D^{\parallel,{\rm iso}}(R,\omega)$ when the firers are isotropic compression centers, as discussed in Sec.~\ref{s:isotropic}.}
\label{f:2pt_coupling_schematic}
\end{figure}

$D_{ij}$ admits a similar scaling form as for $C_{ij}$, but with the inter--particle spacing~$R$ instead of the probe radius $a$, $D_{ij}\sim\bar{c}\kappa^{2}/\mu^{2}R$.
As for $C_{ij}$, the coupling factors $\psi^{\parallel}$, $\psi^{\perp}$ exhibit different high and low frequency behaviors, but now the crossover frequency is determined by the ratio $R/\ell$ instead of $a/\ell$, as schematically shown in Fig.~\ref{f:2pt_coupling_schematic}. The decoupled limit arises at low frequencies such that $\ell\gg R$, when $\psi^{\perp}(z)\rightarrow1$ and $\psi^{\parallel}\rightarrow0$. Conversely in the fully coupled limit $\ell\ll R$, both $\psi^{\perp}$ and $\psi^{\parallel}$ vanish. Note that $\psi^{\parallel}$, $\psi^{\perp}$ are closely related to the response functions $\chi^{\parallel}$, $\chi^{\perp}$ derived in~\cite{MacKintosh2008,Levine2009}.

Despite the dependence of $D_{ij}$ on $R$ through the coupling functions $\psi^{\perp,\parallel}(R/l)$, only the overall scaling $1/R$ has been observed experimentally~\cite{Lau2003,Mizuno2009}. The reason is simply that, for typical actin--myosin parameters, this additional source of variation is weak over the ranges of $R$ typically probed, as confirmed in the inset of Fig.~\ref{f:example}. The overall scaling $R^{-1}$ is a direct consequence of the dipolar nature of the active agents. Dimensional analysis suggests higher--order force multipoles will obey higher powers of $R$, {\em i.e.} for {\em quadrupoles} of the form $f({\bf r})\sim\kappa_{Q}\nabla^{2}\delta({\bf r})$ with $\kappa_{Q}$ the quadrupole moment, $D_{ij}\propto\bar{c}\kappa_{Q}^{2}/\mu^{2}R^{3}$ is expected (see Appendix~\ref{app:Dij}). Therefore the experimental $D_{ij}\sim 1/R$~\cite{Lau2003,Mizuno2009} provides independent confirmation that activated myosin mini--filaments do indeed act as force dipoles to leading order.

In the high frequency regime, or for an incompressible network with $\nu=1/2$ for which $B$ diverges, the coupling terms in (\ref{e:Dij_result}) vanish and $D^{\parallel}=2D^{\perp}$. The same ratio between parallel and perpendicular correlations also holds for thermal fluctuations in incompressible, passive media~\cite{Levine2000}. Although one might hypothesize the existence of universality here, contrary evidence can be obtained by repeating the calculations for the force quadrupoles mentioned above. As described in Appendix~\ref{app:Dij}, in this case we expect $D^{\parallel}=-2D^{\perp}$, so the ratio between $D^{\parallel}$ and $D^{\perp}$ in athermal fluctuation is not solely determined by a material's mechanical properties. In particular, by mixing a population of dipoles and quadrupoles of suitable moments it is hypothetically possible to generate a continuous range of ratios.

%
\subsection{Isotropic compression centers}
\label{s:isotropic}

For the isotropic compression centers given by~(\ref{e:f_iso_dip}), $C_{ij}^{\rm iso}$ and $D_{ij}^{\rm iso}$ can be found by following the same procedure as above, with the propagator ${\bf P}^{\rm iso}({\bf q},\omega)$ in~(\ref{e:propagator}) now defined in terms of the isotropic force dipole from~(\ref{e:f_dip_choice}). Repeating the calculations gives

\begin{eqnarray}
C_{ij}^{\rm iso}(\omega)
&=&
\frac{\bar{c}\kappa^{2}}{108\pi a|B|^{2}}\delta_{ij}\psi(2a/\pi\ell)
\label{e:iso_Cij}
\\
D_{\parallel}^{\rm iso}(\omega)
&=&
\frac{\bar{c}\kappa^{2}}{36\pi R|B|^{2}}
\psi^{\parallel}(R/\ell)
\nonumber\\
D_{\perp}^{{\rm iso}}(\omega)
&=&
\frac{\bar{c}\kappa^{2}}{72\pi R|B|^{2}}
\psi^{\perp}(R/\ell)
\label{e:iso_Dij}
\end{eqnarray}

\noindent{}where the coupling functions $\psi$ are the same as before. Note that the shear terms have now vanished, and thus the variation in frequency due to coupling is far broader than before, suggesting a system of isotropic dipoles would present a stronger test for the prediction of the two--fluid model than the anisotropic case. Both $C_{ij}$ and $D_{ij}$ vanish in the strongly--coupled limit. Since $\psi^{\parallel}(z)\rightarrow0$ as $z\rightarrow0$, the parallel correlations $D^{\parallel}(R,\omega)$ also vanish at {\em low} frequencies, which is most easily explained by reference to the real--space calculations below.

Calculations for an elastic medium in real space are carried out following the same procedure as before, but for the displacement field due to an isotropic dipole. Details are given in Appendix~\ref{app:elasticity_calculations}. The resulting expressions are

\begin{eqnarray}
C_{ij}^{\rm iso,el}
&=&
\left(\frac{1-2\nu}{1-\nu}\right)^{2}
\frac{\bar{c}\kappa^{2}}{432\pi\mu^{2}}
\frac{1}{a}
\delta_{ij}
\label{e:iso_Cij_el}
\\
D_{\parallel}^{\rm iso,el}
&=&
0
\nonumber\\
D_{\perp}^{{\rm iso,el}}
&=&
\left(\frac{1-2\nu}{1-\nu}\right)^{2}
\frac{\bar{c}\kappa^{2}}{288\pi\mu^{2}}
\frac{1}{R}
\label{e:iso_Dij_el}
\end{eqnarray}

\noindent{}The results in all cases (including $C^{\rm iso}_{ij}$) match the low--frequency limits of the expressions~(\ref{e:iso_Cij}) and (\ref{e:iso_Dij}). As in~(\ref{e:iso_Dij}), $D_{\parallel}^{\rm iso}$ vanishes in the low--frequency, elastic limit. This is seen to arise, in real space calculation, from an exact cancellation of firers in between the two probe particles, which generate negative correlations, and those far away, which generate positive correlations; see Fig.~\ref{f:2pt_coupling_schematic}(b). This cancellation requires a displacement field that decays as the inverse--square with distance and would not occur for {\em e.g.} quadrupoles, which instead decay as $1/r^{3}$.

%
%
\section{Discussion}
\label{s:discussion}


The expressions for athermal fluctuations, $C_{ij}(\omega)$ and~$D_{ij}({\bf R},\omega)$ derived in this study serve a number of uses. As explained above, the scaling of $D^{\parallel}$ and $D^{\perp}$ with distance $R$ can be used to infer the leading--order force multipole representative of the firers. In an active gel, myosin minifilaments might work either as oriented dipole or isotropic compression depending on their aggregation and geometries. By comparing the measured parallel and perpendicular correlations, it is also possible to determine the symmetry of the active firers (oriented or isotropic). For materials obeying the 2--fluid model, we can further predict the extent of deviation from the incompressible limits of $C_{ij}$ and $D_{ij}$ in terms of microscopic parameters, as expressed through the coupling functions $\psi$, $\psi^{\parallel}$ and~$\psi^{\perp}$. In actin--myosin systems, this effect is found to be weak, accounting for a $\sim$~10---30\% variation over the usual experimental frequency window, comparable to experimental precision. We can therefore extract the product $\bar{c}|\kappa(\omega)|^{2}/|G(\omega)|^{2}$, to good approximation, by fitting microrheology data to these expressions. Further measuring $G(\omega)$ independently with active microrheology, the quantity $\bar{c}|\kappa(\omega)|^{2}$ can be determined.

Active media are generally expected to have viscoelastic properties different from their equivalent passive systems. For filament networks with active crosslinks ({\em e.g.} myosin mini--filaments), the forces generated by microscopic firers place the filaments in a state of non--zero mean tension, in contrast to a passive network which is on average un\-strained under thermal fluctuations. Semiflexible polymers are known to stiffen under tension, thus an activated filament network is stiffer than a passive one {\em in vitro}~\cite{Mizuno2007}. The local action of force firers $\bar{c}|\kappa(\omega)|^{2}$, which can be quantified using the expressions given in this study, is the microscopic origin of the observed macroscopic active stiffening.  Here we further provide two tentative scaling arguments that lead to expressions for the mean filament tension $\sigma$ in terms of $\bar{c}|\kappa(\omega)|^{2}$, which can be used to predict the macroscopic pre--stressed network response~\cite{Mizuno2007}. One argument is microscopic in nature, the second at the continuum level, and we show at the end are they in broad agreement with each other.

For the microscopic argument, we first need an estimate of the number of active myosin crosslinks per actin filament. Taking the volume associated with a filament of length $\ell$ in a network of mesh size $\xi$ to be $\sim\ell\xi^{2}$, then the typical number of myosin crosslinks per filament should scale as $\sim\bar{c}\ell\xi^{2}$. Substituting typical values for {\em in vitro} actin--myosin systems ($\xi\sim100$nm, $\ell\sim1-10\mu$m and $\bar{c}\sim1\mu$m$^{-3}$~\cite{Mizuno2009b}) gives $<{\mathcal O}(1)$ active crosslinks per filament. We therefore ignore the possibility of multiple active forces on the same filament. The longitudinal component of a force applied to a filament will propagate a distance $\sim\lambda$ before being transferred into bending modes of filaments in the surrounding network, where $\lambda$ is related to the mean distance between crosslinks $\ell_{\rm c}$ (as measured along the filament contour) by $\lambda\sim\ell_{\rm c}(\ell_{\rm c}/d_{\rm A})^{z}$ with $z$ a small exponent and $d_{A}$ the characteristic filament diameter~\cite{Head2003a,Head2003b}. Therefore the mean tension $\sigma$ in the network is estimated to be the number of active crosslinks per filament scaled by $\lambda/\ell$, multiplied by the force per active crosslink $\sim\kappa/\varepsilon$. Assuming that the motor generated forces act between crosslinks~\cite{Mizuno2007}, $\varepsilon\sim\ell_{\rm c}$, the mean tension in the active network is obtained as

\begin{equation}
\sigma
\sim
\bar{c}\kappa\xi^{2}
\left(\frac{\ell_{\rm c}}{d_{\rm A}}\right)^{z}
\quad.
\label{e:tension_micro}
\end{equation}

\noindent{}The exponent $z$ has been predicted to be $2/5$~\cite{Head2003b}, although the slightly different value $z=1/3$ was empirically better in athermal, two--dimensional networks at intermediate crosslinking densities~\cite{Head2003a}.

For the continuum calculation, we consider a single oriented force dipole in an isotropic elastic body, which, when active, generates the strain field $u_{ij}$ derived in Appendix~\ref{app:elasticity_calculations}. The relative extension of a filament segment of length $\ell_{\rm c}$ is estimated as  $\delta\ell_{\rm c}/\ell_{\rm c}=u_{ij}\hat{n}_{i}\hat{n}_{j}$ with $\hat{\bf n}$ the filament orientation~\cite{LandauLifschitz}, assuming $u_{ij}$ is uniform over its length. Averaging $u_{ij}$ over all filament and dipole orientations trivially gives a zero mean extension~$\delta\ell_{\rm c}=0$. However, filaments can only support limited compression $\delta\ell_{\rm c}<0$ before relaxing their longitudinal stress {\em via} buckling. Therefore the integration should be restricted to purely elongating filaments~$\delta\ell_{\rm c}>0$ within a range $X$ from the dipole. At this length $X$, the strain is just large enough to initiate buckling; beyond~$X$, both compression and elongation are allowed and the contribution to $\delta\ell_{c}$ averages to zero. Averaging and integrating $ u_{ij}$ with this restriction, we obtain that the mean extension per filament is $\delta\ell_{\rm c}/\ell_{\rm c}\sim(\bar{c}\kappa/\mu)\ln(X/\ell_{\rm c})$. The logarithmic factor $\ln(X/\ell_{\rm c})$ derives from the inverse--cube decay of the strain field with distance~$r$, $u_{ij}\sim r^{-3}$, necessitating both long and short wavelength cut--offs. Since it is logarithmic, it varies only slowly with $X$ and $\ell_{\rm c}$ and may be regarded as a constant in most cases. To calculate the mean tension~$\sigma$, we use the known force--extension relation $\sigma\sim(\ell_{p}^{2}k_{\rm B}T/\ell_{\rm c}^{3})(\delta\ell_{\rm c}/\ell_{\rm c})$ with $\ell_{\rm p}$ the filament's persistence length~\cite{MacKintosh1995}, giving the final estimate

\begin{equation}
\sigma
\sim
\frac{\ell_{\rm p}^{2}k_{\rm }T}{\ell_{\rm c}^{3}}
\frac{\bar{c}\kappa}{\mu}
\ln(X/\ell_{\rm c})
\quad.
\label{e:tension_cont}
\end{equation}
\noindent{}Regarding the weakly--varying multipliers $\ln(X/\ell_{\rm c})$ and $(\ell_{\rm c}/d_{\rm A})^{z\ll1}$ as numerical prefactors, (\ref{e:tension_micro}) and (\ref{e:tension_cont}) give us the network shear modulus $\mu\sim\ell_{p}^{2}k_{\rm B}T/\ell_{\rm c}^{3}\xi^{2}$ in agreement with prior studies~\cite{MacKintosh1995}. Despite the very different approximations taken, these two estimates are thus broadly consistent. Therefore by determining $\bar{c}|\kappa(\omega)|^{2}$ and $G(\omega)$ using microrheology experiments and the expressions derived in this manuscript, it is now possible to completely address the mechanism of how the nonequilibrium activity leads to the  modification of the network viscoelasticity via (\ref{e:tension_micro}) or~(\ref{e:tension_cont}).

\begin{acknowledgements}
The authors would like to thank F. MacKintosh and C. Schmidt for useful discussions regarding this work. 
\end{acknowledgements}

%
%
\appendix				

%
%
\section{Finite release--time model}
\label{app:finite_release}

Here we derive the power spectrum $|\kappa(\omega)|^{2}$ for the stochastic grow--and--release model given in Fig.~\ref{f:finite_release}. In steady state, the dipole moment $\kappa(t)$ will be growing with probability $p_{\rm G}=\dot{\toy}_{\rm R}/(\dot{\toy}_{\rm G}+\dot{\toy}_{\rm R})$ and releasing with probability $p_{\rm R}=1-p_{\rm G}=\dot{\toy}_{\rm G}/(\dot{\toy}_{\rm G}+\dot{\toy}_{\rm R})$. To find $Q(\toy,t;u_{0})$, the probability density function of $\kappa$ at time $t$ given $\toy=\toy_{0}$ at $t=0$, we therefore integrate over all possible histories and weight according to $p_{\rm G}$ or $p_{\rm R}$. The result is

\begin{eqnarray}
\lefteqn{
Q(\toy,t; \toy_{0})}
\nonumber\\
&=&p_{\rm G}e^{-t/\tau}\delta(\toy-[t\dot{\toy}_{\rm G}+ \toy_{0}])
\nonumber\\
&+&
p_{\rm G}
\theta(\toy-[\toy_{0}-t\dot{\toy}_{\rm R}])
\theta(\toy_{0}+t\dot{\toy}_{\rm G}-\ \toy)
e^{-t/T}
\frac{e^{-\frac{\toy-\toy_{0}}{T\dot{\toy}_{\rm R}}}
}{T\dot{\toy}_{\rm R}}
\nonumber\\
&+&
p_{\rm R}\delta(\toy-[\toy_{0}-t\dot{\toy}_{\rm R}])
\nonumber\\
&+&
p_{\rm R}\theta(t-\toy_{0}/\dot{\toy}_{\rm R})\delta(\toy-\dot{\toy}_{\rm G}[t-\toy_{0}/\dot{\toy}_{\rm R}])
e^{-t/\tau}
e^{\toy_{0}/\tau\dot{\toy}_{\rm R}}
\nonumber\\
&+&
\theta\left(t-\left[\frac{\toy_{0}}{\dot{\toy}_{\rm R}}+\frac{\toy}{\dot{\toy}_{\rm G}}\right]\right)
\times
\nonumber\\
&&
\left\{
\frac{1}{\tau\dot{\toy}_{\rm G}}e^{-\toy/\tau\dot{\toy}_{\rm G}}
-
p_{\rm G}
e^{-t/T}
\frac{1}{T\dot{\toy}_{\rm R}}
e^{-(\toy-\toy_{0})/T\dot{\toy}_{\rm R}}
\right\}
\label{e:P_ut}
\end{eqnarray}

\noindent{}where $T=\tau(\dot{\toy}_{\rm G}+\dot{\toy}_{\rm R})/\dot{\toy}_{\rm R}$ is the mean total cycle time (growth plus release), $\theta$ is the Heaviside step function and $\toy\geq0$ is implicitly assumed throughout. The first term on the right hand side of (\ref{e:P_ut}) corresponds to an initially growing $\toy$ that does not start releasing by time~$t$, the second to those that start releasing at some time $0<s<t$ and are still in this release phase at~$t$, the third to initially releasing~$\toy $, and the fourth and fifth terms to second and subsequent cycles after the first release phase has finished. The derivation of (\ref{e:P_ut}) is straightforward; for instance, the second term is \mbox{$p_{\rm A}\int^{t}_{t_{1}}{\rm d}s\,\frac{1}{\tau}e^{-s/\tau}\delta(\toy-[\toy_{0}+\dot{\toy}_{\rm G}s-(t-s)\dot{\toy}_{\rm R}])$}, where the exponential gives the probability for the first release event at~$s$, the delta function maps to the corresponding value of $\kappa$ at~$t$, and \mbox{$t_{1}=\max(0,[t\dot{\toy}_{R}-\toy_{0}]/[\dot{\toy}_{G}+\dot{\toy}_{R}])$} ensures $\toy\geq0$. Performing the integral gives the second term in~(\ref{e:P_ut}). The fourth and fifth terms are derived in a similar manner. We note that $Q(\toy,t; \toy_{0})$ is properly normalized, $\int_{0}^{\infty}{\rm d} \toy\,Q(\toy,t; \toy_{0})=1$ $\forall\,t$, $\toy_{0}$.

As $t\rightarrow\infty$, $Q(\toy,t; \toy_{0})\rightarrow Q_{\infty}(\toy)=\frac{1}{\tau\dot{\toy}_{\rm G}}e^{-\toy/\tau\dot{\toy}_{\rm G}}$, which is just an exponential distribution with mean $\langle \toy\rangle=\tau\dot{\toy}_{\rm G}$. The correlation function $C(t)=\langle \toy(t_{0}+t) \toy(t_{0})\rangle_{t_{0}}-\langle \toy\rangle^{2}$ in steady state is found by integrating over (\ref{e:P_ut}) suitably weighted by $Q_{\infty}(\toy_{0})$,

\begin{eqnarray}
C(t)&=&
\int_{0}^{\infty}{\rm d}\toy_{0}\,
\int_{0}^{\infty}{\rm d}\toy\,
\toy_{0}\toy\,
Q_{\infty}(\toy_{0})
Q(\toy,t; \toy_{0})-\langle\toy\rangle^{2}
\nonumber\\
&=&
\langle\toy\rangle^{2}
\left\{
\frac{1}{1-\alpha}e^{-t/\tau}
+
\frac{1}{1-\alpha^{-1}}
e^{-t/\alpha\tau}
\right\}
\label{e:corrn}
\end{eqnarray}

\noindent{}using $\alpha=\dot{\toy}_{\rm G}/\dot{\toy}_{\rm R}<1$ for the ratio of growth to release rates. This exhibits positive correlations on the characteristic time scale for growth~$\tau$, and negative correlations on the release time scale $\alpha\tau$. In the rapid release limit $\alpha\rightarrow0$ we recover the expected correlation for a Poisson process, $C(t)=\langle\toy\rangle^{2}e^{-t/\tau}$. Note that although we assume $\alpha<1$ for this discussion and in the main text, the result is also valid for $\alpha>1$ and indeed the steady state is invariant under $(\alpha,\dot{\toy}_{\rm G},\dot{\toy}_{\rm R},\tau)\leftrightarrow(\alpha^{-1},\dot{\toy}_{\rm R},\dot{\toy}_{\rm G},\alpha\tau)$.

It is now straightforward to Fourier transform (\ref{e:corrn}) to get the power spectrum for a single firer $|\kappa(\omega)|^{2}=\int_{-\infty}^{\infty}{\rm d}t\,e^{i\omega t}C(|t|)$, which is given in~(\ref{e:PSD}). For completeness we also give here the limiting case $\dot{\toy}_{\rm G}\rightarrow\dot{\toy}_{\rm R}$ when the cycle is symmetric. We find that $C(t)\sim\langle\toy\rangle^{2}(1+t/\tau)e^{-t/\tau}$ and $|\kappa(\omega)|^{2}=4\tau\langle\toy\rangle^{2}/(1+\omega^{2}\tau^{2})^{2}$, so the pseudo--diffusive regime is now entirely absent.

%
%
\section{Poisson ratio for affine network deformation}
\label{app:poisson_ratio}

Here we derive the Poisson ratio for a 3--dimensional filament network under the assumptions of affine quasi--static deformation and a homogeneous, isotropic network. Consider a single filament of length $\ell$ and orientation $\hat{\bf n}$ with respect to some fixed axes. If the system is deformed to give a displacement field ${\bf u}({\bf x},t)$, and assuming this macroscopic field scales uniformly down to lengths $\ll\ell$ ({\em i.e.} the deformation is affine), then the filament elongates by an amount $\delta\ell$ given by~\cite{LandauLifschitz}

\begin{equation}
\frac{\delta\ell}{\ell}
=
u_{ij}\hat{n}_{i}\hat{n}_{j}
\end{equation}

\noindent{}where $u_{ij}=(\partial_{i}u_{j}+\partial_{j}u_{i})/2$ is the linear strain tensor. The increase in energy in this filament's deformed state is then $\delta E=\frac{1}{2}k(\delta\ell)^{2}=\frac{1}{2}k\ell^{2}(u_{ij}\hat{n}_{i}\hat{n}_{j})^{2}$ with $k$ an effective spring constant. Note that $k$ may depend on $\ell$, the frequency of deformation change, any pre--tensile stress in the filament~{\em etc}. For the purposes of this calculation, however, we need to only assume $k$ is independent of the filament orientation~$\hat{\bf n}$.

Now consider a homogeneous population of filaments with length distribution ${\mathcal L}(\ell)$, orientation distribution ${\mathcal N}(\hat{\bf n})$ and number density~$\rho$. To ensure isotropy, $\ell$ and $\hat{\bf n}$ are uncorrelated and ${\mathcal N}(\hat{\bf n})=(4\pi)^{-1}$. The increase in energy per unit volume is then

\begin{eqnarray}
\Delta E
&=&
\rho
\int{\rm d}\ell\,
{\mathcal L}(\ell)
\int{\rm d}\hat{\bf n}\,
{\mathcal N}(\hat{\bf n})
\frac{1}{2}
k\ell^{2}
(u_{ij}\hat{n}_{i}\hat{n}_{j})^{2}
\\
&=&
\Sigma
\frac{1}{4\pi}\int{\rm d}\hat{\bf n}\,
(u_{ij}\hat{n}_{i}\hat{n}_{j})^{2}
\end{eqnarray}

\noindent{}where $\Sigma=\frac{1}{2}\rho\int{\rm d}\ell{\mathcal L}(\ell)k\ell^{2}$. Assuming quasi--static response, this energy change equals the corresponding elastic energy $\Delta E^{\rm el}=\frac{1}{2}\lambda u_{ii}^{2}+\mu u_{ij}^{2}$ with $\lambda$, $\mu$ the Lam\'e coefficients~\cite{LandauLifschitz}. By choosing two independent strain fields $u_{ij}$ it is therefore possible to determine both $\lambda$ and~$\mu$. For instance, an isotropic compression $u_{ij}=-\gamma\delta_{ij}$ gives the equation $\gamma^{2}\Sigma=\frac{9}{2} \lambda\gamma^{2}+3\mu\gamma^{2}$. For simple shear, where $u_{xy}=u_{yx}=\gamma/2$ and $u_{ij}=0$ otherwise, another equation $\gamma^{2}\Sigma/15=\mu\gamma^{2}/2$ is obtained. From these 2 equations, we obtain $\lambda=\mu=2\Sigma/15$, and the Poisson ratio $\nu=\lambda/[2(\lambda+\mu)]=1/4$, independent of $\Sigma$ and hence~$k$. By coincidence, this is in good agreement with the value obtained for semi--dilute flexible polymers in a good solvent~\cite{Geissler1980,Geissler1981}.

%
%
\section{Real-space calculations}
\label{app:elasticity_calculations}

The displacement field ${\bf u}({\bf r})$ due to a point force ${\bf f}$ applied at the origin of an infinite, homogeneous, isotropic elastic continuum is known, \mbox{$u_{i}({\bf r})=[16\pi\mu(1-\nu)r]^{-1}\left\{(3-4\nu)\delta_{ij}+{\hat r}_{i}{\hat r}_{j}\right\}f_{j}$}, with $\mu$ the shear modulus, $\nu$ the Poisson ratio, $r=|{\bf r}|$ and $\hat{\bf r}={\bf r}/r$~\cite{LandauLifschitz}. The displacement field extending from the dipole with orientation $\hat{\bf n}$ defined in~(\ref{e:f_dip}) is then obtained by taking the gradient of the monopole solution,

\begin{eqnarray}
u^{\rm ani}_{i}({\bf r};\hat{\bf n})
&=&
\frac{\kappa}{16\pi\mu(1-\nu)}\frac{1}{r^{2}}
\Big\{
2(1-2\nu)[\hat{\bf r}\cdot\hat{\bf n}]\hat{n}_{i}
\nonumber\\
&&
-(1-3[\hat{\bf r}\cdot\hat{\bf n}]^{2})\hat{r}_{i}
\Big\}
\end{eqnarray}

\noindent{}with $\kappa=\varepsilon f$ the dipole moment. The linear strain $u_{ij}=(\partial_{i}u_{j}+\partial_{j}u_{i})/2$ and stress fields can be found by the usual relations. For instance, the strain is

\begin{eqnarray}
\lefteqn{
u_{ij}^{\rm ani}({\bf r};\hat{\bf n})
=
-\frac{\kappa}{16\pi\mu(1-\nu)}
\frac{1}{r^{3}}
\times
}
\nonumber\\
&&\Big{\{}
2(1-2\nu)\hat{n}_{i}\hat{n}_{j}
-
3(5[\hat{\bf r}\cdot\hat{\bf n}]^{2}-1)\hat{r}_{i}\hat{r}_{j}
\nonumber\\
&&
+
(3[\hat{\bf r}\cdot\hat{\bf n}]^{2}-1)\delta_{ij}
+
6\nu[\hat{\bf r}\cdot\hat{\bf n}]
(\hat{n}_{i}\hat{r}_{j} + \hat{n}_{j}\hat{r}_{i})
\Big{\}}.
\label{e:el_ani_strain}
\end{eqnarray}

\noindent{}This equation has a trace $u^{\rm ani}_{ii}=-\kappa(1-2\nu)[8\pi\mu(1-\nu)r^{3}]^{-1}(1-3[\hat{\bf r}\cdot\hat{\bf n}]^{2})$ that vanishes in the incompressible limit $\nu=1/2$ as expected.

It is possible to derive an exact solution for finite dipole with size~$\varepsilon$ by exploiting the spherical symmetry that arises after averaging over~$\hat{\bf n}$. The resultant displacement field is found by integrating the monopole solution over a sphere of inward--pointing monopole forces $-2f\hat{\bf r}$. The sphere has radius $\varepsilon/2$ centered on the origin, and the factor 2 arises since each dipole consists of 2 monopoles. By symmetry only the response at, say, a point $(0,0,z)$ needs to be considered. The integration is simplified by switching to spherical polar coordinates and making the change of variables $u=\sqrt{r^{2}+\varepsilon^{2}/4-r\varepsilon\cos\theta}$ for the polar angle $\theta$; the cases $z<\varepsilon/2$ and $z>\varepsilon/2$ correspond to different integral limits and must be considered separately. The results are presented in Table~\ref{t:isotropic}.

\begin{table}
\caption{The displacement, strain and stress fields for an isotropic compression center of size $\varepsilon$ centered on the origin. $S_{ij}$ is the traceless tensor $S_{ij}=\frac{1}{3}\delta_{ij}-\hat{r}_{i}\hat{r}_{j}$.
\label{t:isotropic}}
\begin{center}
\begin{tabular}{l|c|c}
Field & $r<\varepsilon/2$ & $r>\varepsilon/2$
\\
\hline
\hline
Displacement $u^{\rm iso}_{i}({\bf r})$
&
$\displaystyle{-\frac{\kappa(1-2\nu)}{3\pi\mu(1-\nu)}
\frac{r}{\varepsilon^{3}}\hat{r}_{i}}$
&
$\displaystyle{-\frac{\kappa(1-2\nu)}{24\pi\mu(1-\nu)}
\frac{1}{r^{2}}\hat{r}_{i}}$
\\
\hline
Strain $u^{\rm iso}_{ij}({\bf r})$
&
$\displaystyle{-\frac{\kappa(1-2\nu)}{3\pi\mu(1-\nu)}\frac{1}{\varepsilon^{3}}\delta_{ij}}$
&
$\displaystyle{-\frac{\kappa(1-2\nu)}{8\pi\mu(1-\nu)}\frac{1}{r^{3}}}S_{ij}$
\\
\hline
Trace $u^{\rm iso}_{ii}({\bf r})$
&
$\displaystyle{-\frac{\kappa(1-2\nu)}{\pi\mu(1-\nu)}
\frac{1}{\varepsilon^{3}}}$
&
0
\\
\hline
Stress $\sigma^{\rm iso}_{ij}({\bf r})$
&
$\displaystyle{-\frac{2\kappa(1+\nu)}{3\pi(1-\nu)}
\frac{1}{\varepsilon^{3}}\delta_{ij}}$
&
$\displaystyle{-\frac{\kappa(1-2\nu)}{4\pi(1-\nu)}\frac{1}{r^{3}}}S_{ij}$
\end{tabular}
\end{center}
\end{table}

The firers are assumed to be independently and uniformly distributed throughout space, so the firer concentration field $c({\bf x})$ obeys $\langle c({\bf x})c({\bf y})\rangle=\bar{c}\delta({\bf x}-{\bf y})$ and hence

\begin{eqnarray}
\lefteqn{\langle u_{i}({\bf 0})u_{j}({\bf R})\rangle}
\nonumber\\
&=&
\left\langle
\int{\rm d}{\bf x}\int{\rm d}{\bf y}\,c({\bf x})c({\bf y})u_{i}(-{\bf x})u_{j}({\bf R}-{\bf y})
\right\rangle
\\
&=&\bar{c}\int{\rm d}{\bf x}\,u_{i}(-{\bf x})u_{j}({\bf R}-{\bf x})
\end{eqnarray}

\noindent{}The single--point correlations correspond to~${\bf R}={\bf 0}$. The required expression for a single isotropic or anisotropic field $u_{i}$ is then inserted into the integrand. For the 2--point correlations it is convenient to set ${\bf R}=(0,0,R)$ for the integration, and rotate to general $\hat{\bf R}$ at the end. The integration is facilitated by switching to spherical polar coordinates by converting variables as $u=\sqrt{r^{2}+R^{2}-2rR\cos\theta}$. We give here an intermediate expression for the correlations due to a nematic field of anisotropic dipoles with the same orientation $\hat{\bf n}$,

\begin{eqnarray}
\lefteqn{
D^{\rm ani}_{ij}
=
\frac{\bar{c}\kappa^{2}}{256\pi\mu^{2}(1-\nu)^{2}R}
(1-[\hat{\bf n}\cdot\hat{\bf R}]^{2})
\times}
&&
\nonumber\\
&&
\left\{
(1-[\hat{\bf n}\cdot\hat{\bf R}]^{2})\delta_{ij}
+
4[\hat{\bf n}\cdot\hat{\bf R}](2-3\nu)
[\hat{n}_{i}\hat{R}_{j}+\hat{R}_{i}\hat{n}_{j}]
\right.
\nonumber\\
&&
\left.
+4(1-2\nu)(3-4\nu)\hat{n}_{i}\hat{n}_{j}
-
(1-5[\hat{\bf n}\cdot\hat{\bf R}]^{2})
\hat{R}_{i}\hat{R}_{j}
\right\}
\nonumber\\
\label{e:nematic}
\end{eqnarray}

\noindent{}The expression given in (\ref{e:Dij_elastic}) is then found by averaging (C5) over~$\hat{\bf n}$. For isotropic dipoles we assume $\varepsilon/r\rightarrow0$ and hence only use the $r>\varepsilon/2$ expressions in Table~\ref{t:isotropic}.

%
%
\section{$D_{ij}$ for the 2--fluid model}
\label{app:Dij}

$D_{ij}({\bf q},\omega)$ for the 2--fluid model is found by inserting the anisotropic dipole propagator ${\bf P}({\bf q},\omega)$ from~(\ref{e:propagator}) and~(\ref{e:f_dip_choice}) into~(\ref{e:Dij_defn}) and then averaging over  orientation~$\hat{\bf n}$. In terms of projection operators parallel and perpendicular to $\hat{\bf q}$,

\begin{eqnarray}
D_{ij}({\bf q},\omega)
&=&
E^{\parallel}(q)\hat{q}_{i}\hat{q}_{j}
+
E^{\perp}(q)(\delta_{ij}-\hat{q}_{i}\hat{q}_{j})
\nonumber\\
E^{\parallel}(q)
&=&
\bar{c}\kappa^{2}
\frac{q^{2}}{5|Bq^{2}-i\omega\Gamma|^{2}}
\nonumber\\
E^{\perp}(q)
&=&
\bar{c}\kappa^{2}
\frac{1}{15q^{2}|G|^{2}}
\label{e:reciprocal_Dij}
\end{eqnarray}

\noindent{}Note this projection is in ${\bf q}$--space, not real space. For an isotropic material $D_{ij}({\bf R})$ can similarly be projected into components parallel and perpendicular to the real--space unit vector $\hat{\bf R}$, {\em i.e.} \mbox{$D_{ij}({\bf R})=D^{\parallel}(R)\hat{R}_{i}\hat{R}_{j}+D^{\perp}(R)(\delta_{ij}-\hat{R}_{i}\hat{R}_{j})$}.

To relate the ${\bf q}$--space projections to the real space ones, note that \mbox{$\hat{R}_{i}\hat{R}_{j}D_{ij}({\bf R})=D^{\parallel}(R)$} and \mbox{$(\delta_{ij}-\hat{R}_{i}\hat{R}_{j})D_{ij}({\bf R})=2D^{\perp}(R)$}. By writing \mbox{$D_{ij}({\bf R})$} as the inverse transform of (\ref{e:reciprocal_Dij}) in both of these expressions, it is readily seen that the following transform pairs hold between $D^{\parallel,\perp}(R)$ and $E^{\parallel,\perp}(q)$,

\begin{eqnarray}
D^{\parallel}(R)
&\stackrel{ {\rm FT} }{\longrightarrow}&
E^{\perp}(q)(1-[\hat{\bf q}\cdot\hat{\bf R}]^{2})
+
E^{\parallel}(q)[\hat{\bf q}\cdot\hat{\bf R}]^{2}
\nonumber\\
2D^{\perp}(R)
&\stackrel{ {\rm FT} }{\longrightarrow}&
E^{\perp}(q)(1+[\hat{\bf q}\cdot\hat{\bf R}]^{2})
+
E^{\parallel}(q)(1-[\hat{\bf q}\cdot\hat{\bf R}]^{2})
\nonumber\\
\label{e:proj2proj}
\end{eqnarray}

\noindent{}Performing the inverse transforms of the $E^{\parallel,\perp}(q)$ in~(\ref{e:reciprocal_Dij}) using~(\ref{e:proj2proj}) gives the final expressions~(\ref{e:Dij_result}). The integrations are facilitated by switching to cylindrical polar coordinates $(\rho,\theta,z)$ and exploiting isotropy to set ${\bf R}=(0,0,R)$ without loss of generality.

It is possible to make some general remarks on the correlations of displacement fields driven by force quadrupoles, $f({\bf r})\sim\kappa_{Q}\nabla^{2}\delta({\bf r})$ with $\kappa_{Q}$ the quadrupole moment, without going through detailed calculations. After averaging over the quadrupole internal degrees of freedom, $D_{ij}$ is expected to take the same form as (\ref{e:reciprocal_Dij}) but with $E^{\perp}(q)\propto \bar{c}\kappa^{2}_{Q}/|G|^{2}$, the different power of $q$ (with respect to dipoles) being due to the one--higher derivative of $\delta({\bf r})$ in~${\bf f}$. The longitudinal component $E^{\parallel}\equiv0$ for an incompressible body. Repeating the inverse transformations (\ref{e:proj2proj}) now gives $D^{\parallel}=-2D^{\perp}\propto\bar{c}\kappa^{2}_{Q}/|G|^{2}R^{3}$, as discussed in the text.


\end{document}